\documentclass{article}

\usepackage{microtype}
\usepackage{graphicx}
\usepackage{subcaption}
\usepackage{booktabs} % for professional tables
\usepackage{multirow}
\usepackage{hyperref}

\usepackage{pifont}% http://ctan.org/pkg/pifont

\usepackage{amsfonts}
\usepackage{amsmath}
\usepackage{amssymb}

\usepackage{svg}

\usepackage[accepted]{icml2021}

\newcommand{\ourmodel}{SFM-Protein}

\begin{document}
    \twocolumn[
    \icmltitle{\ourmodel{}: Integrative Co-evolutionary Pre-training for Advanced Protein Sequence Representation}
    
	% It is OKAY to include author information, even for blind
	% submissions: the style file will automatically remove it for you
	% unless you've provided the [accepted] option to the icml2021
	% package.
	
	% List of affiliations: The first argument should be a (short)
	% identifier you will use later to specify author affiliations
	% Academic affiliations should list Department, University, City, Region, Country
	% Industry affiliations should list Company, City, Region, Country
	
	% You can specify symbols, otherwise they are numbered in order.
	% Ideally, you should not use this facility. Affiliations will be numbered
	% in order of appearance and this is the preferred way.
	%	\icmlsetsymbol{equal}{*}

    \centering
    \textbf{Liang He$^{*, 1}$, Peiran Jin$^{*, 1}$, Yaosen Min$^{*, 1}$, Shufang Xie$^{*, 1}$, Lijun Wu$^{*, 1}$, Tao Qin$^{*, 1}$ 
    \\ Xiaozhuan Liang$^{2}$, Kaiyuan Gao$^{3}$, Yuliang Jiang$^{4}$, Tie-Yan Liu$^{1}$} \\
    \vspace{1em}
    % $^*$Contributed equally to this work \\ % $^\dagger$Corresponding authors \\
    \vspace{1em}
    \textit{$^{1}$\{lihe, peiranjin, yaosenmin, lijuwu, shufxi, taoqin, 
    tyliu\}@microsoft.com} \\
    \textit{$^{2}$liangxiaozhuan@zju.edu.cn}, \
    \textit{$^{3}$im\_kai@hust.edu.cn, \ $^{4}$jiangyl22@mails.tsinghua.edu.cn} \\
    \textit{$^{*}$Equally contributed} \\
    
    % \icmlcorrespondingauthor{Liang He}{lihe@microsoft.com} 
    % \icmlcorrespondingauthor{Tao Qin}{taoqin@microsoft.com} 
    % \\
    \vspace{1em}

	%%% ---------------------------------------------------

    \text{$^{1}$Microsoft Research AI for Science}, \ \text{$^{2}$Zhejiang University} \\
    \text{$^{3}$Huazhong University of Science and Technology}, \ 
    \text{$^{4}$Tsinghua University} \\

	% You may provide any keywords that you
	% find helpful for describing your paper; these are used to populate
	% the "keywords" metadata in the PDF but will not be shown in the document
	% \icmlkeywords{Machine Learning, ICML}
	
	\vskip 0.3in
	]
	%%% TODO: author list here!
	%	\let\thefootnote\relax\footnotetext{$^\dagger$This work was conducted at Microsoft Research AI4Science}
	%	\let\thefootnote\relax\footnotetext{$^\ddagger$Corresponding author}
	
	% this must go after the closing bracket ] following \twocolumn[ ...
	
	% This command actually creates the footnote in the first column
	% listing the affiliations and the copyright notice.
	% The command takes one argument, which is text to display at the start of the footnote.
	% The \icmlEqualContribution command is standard text for equal contribution.
	% Remove it (just {}) if you do not need this facility.
	
	%	\printAffiliationsAndNotice{}  % leave blank if no need to mention equal contribution
	%	\printAffiliationsAndNotice{\icmlEqualContribution} % otherwise use the standard text.

    \begin{abstract}
    Proteins, essential to biological systems, perform functions intricately linked to their three-dimensional structures. Understanding the relationship between protein structures and their amino acid sequences remains a core challenge in protein modeling. While traditional protein foundation models benefit from pre-training on vast unlabeled datasets, they often struggle to capture critical co-evolutionary information, which evolutionary-based methods excel at.
    In this study, we introduce a novel pre-training strategy for protein foundation models that emphasizes the interactions among amino acid residues to enhance the extraction of both short- and long-range co-evolutionary features from sequence data. Trained on a large-scale protein sequence dataset, our model demonstrates superior generalization ability, outperforming established baselines of similar size, including the ESM model, across diverse downstream tasks. Experimental results confirm the model’s effectiveness in integrating co-evolutionary information, marking a significant step forward in protein sequence-based modeling.
    \end{abstract}
	
    \section{Introduction}

Proteins are the molecular machines of life, orchestrating essential biological functions that drive survival and growth. Their functionality stems from complex three-dimensional structures, which are determined by the unique sequences of amino acids that make up each protein~\cite{jumper2021highly,yang2020improved}. Understanding how amino acid sequences define protein structure, and in turn function, remains a key challenge in biology. Solving this puzzle is crucial for designing targeted interventions in diseases and engineering proteins with new, desirable functions~\cite{whitford2013proteins}.

Proteins are far more than simple chains of amino acids. Their behavior cannot be fully understood by examining individual residues or their linear arrangement. Instead, proteins fold into intricate tertiary structures where residues far apart in sequence can be spatially adjacent. These interactions, both short- and long-range, shape the protein's functional roles, mediating both static and dynamic biological processes~\cite{sinai2017variational,riesselman2017deep,ding2019deciphering,hsu2021combining}.

Co-evolutionary interactions, both within local sequences and across distant residues, are critical for maintaining the structural and functional integrity of proteins. These interactions ensure that amino acid co-variations preserve the stability of the protein’s structure. This co-evolution reflects the collective interactions that underpin protein dynamics, influencing their folding and function.
Following Anfinsen's landmark discovery that protein sequences inherently contain the information required to fold into their native structures~\cite{anfinsen1961kinetics}, Levinthal proposed his famous paradox. He argued that the number of possible protein conformations is so vast that a random search would take an impractical amount of time to find the correct structure~\cite{levinthal1968there,levinthal1969fold}. The paradox hinted at the existence of predetermined folding pathways, though their mechanisms remained unclear. To address this, researchers proposed the concept of ``foldons'', small cooperative folding units consisting of multiple residues, which fold sequentially and guide the protein to its final structure~\cite{englander2014nature}. In silico studies have revealed that the formation of local structures often precedes the establishment of distant, non-local contacts~\cite{lindorff2011fast}. These early-forming interactions are key to efficiently achieving the protein’s final, functional form without the need for backtracking~\cite{voelz2007exploring}.
Understanding these co-evolutionary interactions, both short-range, which influence secondary structures, and long-range, which define tertiary structures, is vital for decoding the sequence-structure-function relationship. Insights into this relationship are crucial for applications such as drug design and protein engineering~\cite{phillips2009scaling,phillips2009scaling2,tang2020long,xu2021long}.

Protein foundation models (PFMs), particularly those pre-trained on vast datasets of unlabeled sequences, have emerged as powerful tools for representing protein sequences~\cite{alley2019unified,bepler2019learning,rives2021biological,madani2020progen}. These models achieve remarkable accuracy in protein-related tasks through fine-tuning. However, while models based on multiple sequence alignments (MSAs) effectively capture co-evolutionary relationships between homologous proteins, single sequence-based models often fall short in this area~\cite{remmert2012hhblits,mirabello2019rawmsa}. Despite utilizing the same sequence data, single sequence models underperform compared to MSA-based approaches. We hypothesize that this performance gap arises from the insufficient capture of residue interactions in conventional pre-training techniques, which MSAs handle more effectively by incorporating evolutionary information.

\begin{figure}[!h]
    \centering
    \includegraphics[width=0.35\textwidth]{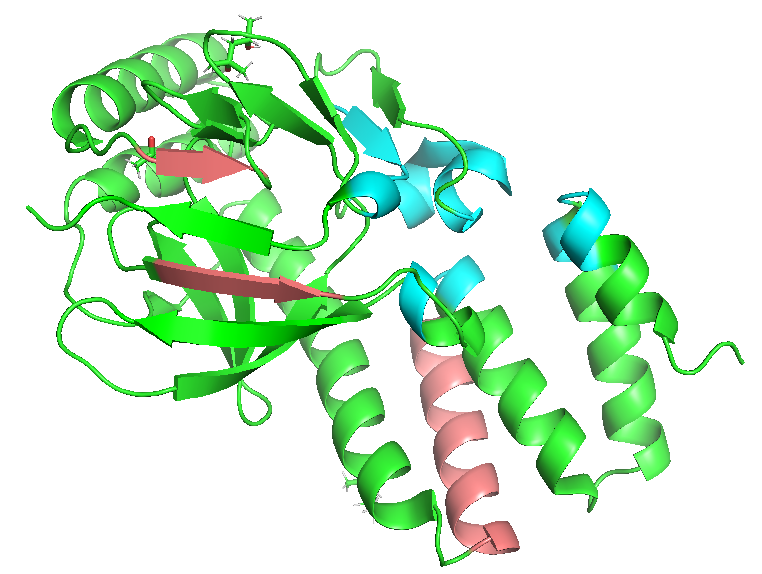}
    \caption{Illustration of Local and Global Interactions. The local interactions are highlighted in red, while the global interactions are highlighted in cyan. (PDB ID: 4UQX)}
    \label{fig:co-evolutionary}
\end{figure}

Inspired by the folding hypothesis and existing research, we categorize co-evolutionary interactions into two types: short-range (local) interactions, which are crucial for forming secondary structures, and long-range (global) interactions, which shape the protein’s overall tertiary structure. This distinction is illustrated in Figure~\ref{fig:co-evolutionary}, where local interactions are highlighted in red and global interactions in cyan.

This paper introduces a novel framework for protein foundation models that effectively captures both local and global co-evolutionary information within protein sequences. Leveraging large protein sequence databases, our model employs an integrative loss function designed to better reflect the complex interplay of interactions that govern protein folding and function. We detail the model's architecture, pre-training strategies, and the innovative loss functions used to capture both short- and long-range co-evolutionary information.
Through extensive validation against empirical data and benchmarking against existing models, our approach demonstrates superior performance in predicting protein properties and functions.

\begin{figure*}[!ht]
    \centering
    \includegraphics[width=0.65\textwidth]{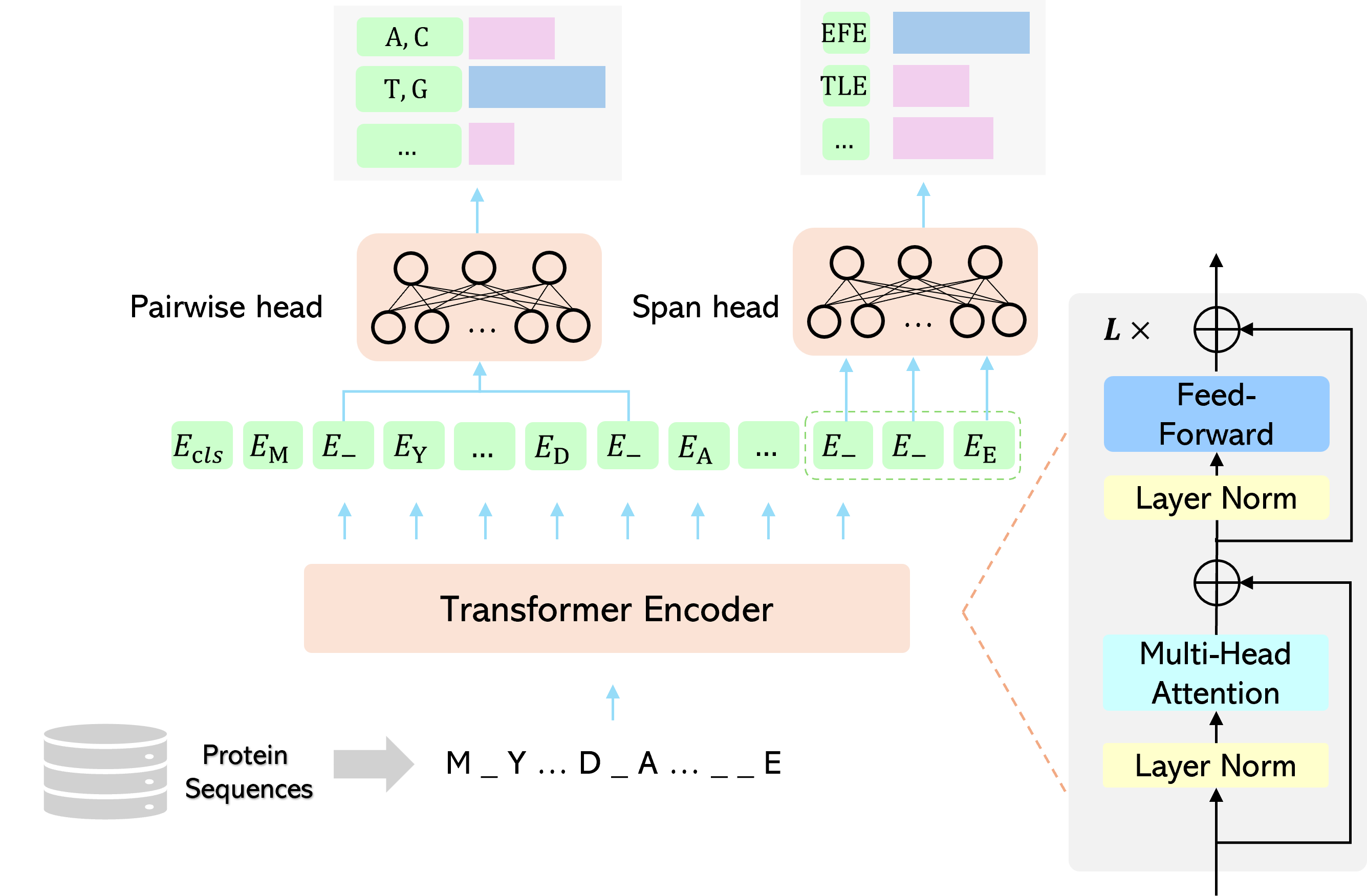}
    % \includesvg[width=0.9\textwidth,inkscapelatex=false]{figures/overview.svg}
    \caption{An overview of the \ourmodel{} framework}
    \label{fig:overview}
\end{figure*}

The remainder of this paper is structured as follows:
Section~\ref{sec:related} reviews related work on protein foundation models.
Section~\ref{sec:methods} outlines our model’s architecture, loss function, and training algorithm.
Section~\ref{sec:exp} presents our experimental results across a variety of downstream tasks.
Section~\ref{sec:discussion} concludes with a discussion of future directions and potential extensions of this work.
    \section{Related Work}
\label{sec:related}

Protein representation learning has become a pivotal area of research within computational biology, with approaches generally classified into two main categories: sequence-based and structure-based methods.

\noindent \textbf{Sequence-based Approaches.}
Traditional sequence-based methods primarily rely on sequence alignment to capture evolutionary information. Tools like PSI-BLAST~\cite{altschul1997gapped} and HMMER~\cite{finn2011hmmer} have been instrumental in identifying evolutionary relationships between protein sequences, which aid in predicting protein structure and function. With the advent of deep learning, protein language models (PLMs) have emerged, leveraging neural networks to encode protein sequences. For example, Bepler et al.~\cite{bepler2019learning} used a BiLSTM to learn protein representations that capture sequential similarity and function. Inspired by advances in natural language processing, Rives et al. introduced ESM-1b~\cite{rives2021biological}, which employed self-supervised learning to create general-purpose sequence representations for downstream tasks. Subsequent models in the ESM series, such as ESM-1v~\cite{meier2021language}, ESM-IF1~\cite{hsu2022learning}, and ESM-2~\cite{lin2022language}, have progressively improved the performance of PLMs on large-scale protein sequence data.
Several methods have integrated multiple sequence alignment (MSA) data to enhance PLMs' representation capacity. Examples include ESM-MSA-1b~\cite{rao2021msa} and Tranception~\cite{notin2022tranception}, which fuse evolutionary information to improve predictions. More recently, models like ESMFold~\cite{lin2023evolutionary}, AminoBERT~\cite{chowdhury2022single}, and OmegaFold~\cite{wu2022high} have explored the potential of single-sequence models for protein structure prediction, showing promising results in bypassing the need for MSA.

\noindent \textbf{Structure-based Approaches.}
With the success of advanced protein structure prediction models, such as AlphaFold~\cite{varadi2024alphafold}, and the availability of large structure databases, there has been increasing interest in developing PLMs that incorporate structural information. One approach encodes protein structures as graphs, where atoms or residues are represented as nodes, and structural contacts define the edges. Graph neural networks (GNNs), such as those used by Hermosilla et al.\cite{hermosilla2022contrastive}, Zhang et al.\cite{zhang2022protein}, and others~\cite{van2024fast, yang2023masked}, have shown promise in capturing geometric structure for downstream tasks. However, GNN-based methods face scalability challenges compared to the transformer-based architectures used in large language models.
Another emerging direction involves tokenizing structural information and integrating it into PLMs. Approaches like UniMol~\cite{Zhou2023UniMolAU}, SAPROT~\cite{su2023saprot}, and structural pre-training methods by Chen et al.~\cite{chen2023structure} have fused structural data with sequence-based models, achieving improved performance in tasks requiring structural understanding. Notably, the recently introduced ESM-3 model~\cite{hayes2024simulating}, designed specifically for protein design, incorporates structural information directly as input, which differentiates it from purely sequence-based models like ours.
Our work distinguishes itself by developing a novel pre-training framework for protein foundation models that enhances the capture of both local and global co-evolutionary information from sequences, without requiring MSA or explicit structural input. This approach aims to close the gap between single-sequence models and MSA-based methods by better modeling residue interactions and co-evolutionary signals through a dedicated loss function tailored to protein folding and structure prediction.
    \section{Methods}
\label{sec:methods}

\ourmodel{} is a protein sequence representation model based on Transformer encoder~\cite{vaswani2017attention}, which leverages bi-directional self-attention to capture the interactions among the residues pre-trained using a masked language model with both pairwise loss and span loss. The model details will be introduced in the following section. 

\subsection{Preliminaries}

A Transformer encoder layer consists of a multi-headed self-attention mechanism and a feed-forward network, both incorporating skip connections and pre-layer normalization~\cite{xiong2020layer}.:

% $$x = \textsc{LayerNorm}(x + \textsc{MHA}(x)), $$
% $$x = \textsc{LayerNorm}(x + \textsc{FFN}(x)), $$

$$x = x + \textsc{MHA}(\textsc{LayerNorm}(x)), $$
$$x = x + \textsc{FFN}(\textsc{LayerNorm}(x)), $$

The multi-head attention can be written as:
$$\textsc{MHA}(Q, K, V) = \textsc{Concat}({head}_1, ..., {head}_h) \, W^O$$
$${head}_i = \textsc{Attn}(Q W^Q_i, K W^K_i, V W^V_i)$$
$$\textsc{Attn}(Q, K, V) = \textsc{Softmax}(\frac{Q K^{T}}{\sqrt{d_k}}) \cdot V$$ 

% While the feed-forward network can be formulated as:
% $$\textsc{FFN}(x) = \sigma(x \, W_1 + b_1) \, W_2 + b_2$$

\ourmodel{} also exploits a positional embedding to distinguish the tokens in different positions, i.e., RoPE (Rotary Positional Embedding), which employs sinusoidal functions to allocate positional values to tokens. Due to the inherent repetitive nature of sinusoidal functions, certain positional values may closely resemble others.
The formal calculation of the embedding is defined by:
$$g(x_m, \, x_n, \, m-n) = Re[ (W_q \cdot x_m) \, (W_k \cdot x_n) \, e^{i(m-n)\theta} ]$$
where $m$ and $n$ are two residue indices within $[0, L]$, $Re$ is the real part of the complex, $W_q, W_k$ are the weight matrices.
This positional embedding is added to the attention during training.

\subsection{Sequence-based Pre-training}

The initial phase of our model training involves pre-training on protein sequences. This pre-training is done in a self-supervised fashion using a unique masked language modeling objective. The model has two output heads: a pairwise prediction head which is used to model pairwise interactions between residues, and a span prediction head which predicts the byte-pair encoding (BPE) tokens of the protein sequences.

Our model's backbone is a deep transformer encoder architecture that is made up of several layers of multi-head self-attention (MHA) and feed-forward networks. To infuse positional information into the self-attention layers, we use relative positional encodings (RoPE).

In our approach, the masked language modeling objective is used to randomly mask a subset of residues for prediction. Notably, instead of masking residues independently, we mask consequential spans of residues with a certain probability $x\%$. This unique span-based masking strategy is designed to encourage the model to learn higher-order residue interactions and patterns. Furthermore, it aims to model the secondary structure in the proteins, which is a significant aspect of understanding and predicting protein function and structure.

\subsection{Global Co-evolution: Pair Label Recovering}

The pairwise prediction head utilizes the output embedding from the transformer to predict a distribution over all possible pairs of residues for each residue in the sequence. This output is then transformed into a pair-wise feature representation as defined by the equation:
$$\mathcal{F}_{global}(x) = W_o * ((W_q \cdot x) \otimes (W_k \cdot x))$$
where $\otimes$ stands for the outer product, i.e., $[x \otimes y]_{ij} = x_i \cdot y_j$, and $\cdot$ is the element-wise product of two vectors.

Following this transformation, the pair-wise feature is then projected into a pair-wise dictionary with the help of a Multilayer Perceptron (MLP). The primary aim of this pairwise recovering task is to enable the model to capture and understand the interactions between different residues in the protein sequence. 

\subsection{Local Co-evolution: BPE Label Recovering}

The span prediction head operates by predicting the BPE tokens based on the contextual representations derived from the model. A BPE tokenizer, trained on the protein sequences from the training data, is used to tokenize each protein sequence into BPE tokens.

Then, each residue is assigned a BPE label that corresponds to the BPE token it is grouped into. If the tokenizer encodes a consecutive span of residues $(x_i, x_{i+1}, ..., x_{i+k})$ into a single BPE token $b$, then all these residue positions are assigned the same BPE label $b$: $x_i = b$, $x_{i+1} = b$, \ldots, $x_{i+k} = b$.

This labeling methodology serves a dual purpose. Firstly, it allows the model to learn representations at the individual residue level via the pairwise prediction head. This allows the model to understand the interactions between different residues. Secondly, it enables learning at the BPE token level through the span prediction head, which helps the model to identify and understand patterns across multiple consecutive residues, providing a more comprehensive understanding of the protein sequences. This dual-level learning is a key differentiator of our model, making it uniquely positioned to capture both local and global contextual information in protein sequences.

\subsection{Training Objectives}

To effectively encapsulate the co-evolutionary information within a comprehensive framework, we propose a composite loss function:

$$\mathcal{L} = \alpha \cdot \mathcal{L}_{global} + (1 - \alpha) \cdot \mathcal{L}_{local}$$

This function is composed of two distinctive components. The global loss, also known as pairwise loss, is specifically designed to capture the co-evolutionary relationships between masked pairs within the sequence. It does this by calculating a cumulative loss across these pairs, with its influence on the overall loss function dictated by the coefficient $\alpha$:

$$\mathcal{L}_{global} = \sum_{X \in \mathcal{D}} \ \sum_{i, j \in M} \log P(x_i, \ x_j \ | \ X_{/M})$$

On the other hand, the local loss, also referred to as span loss, is employed to account for the contiguous masked spans within the sequence. The Byte Pair Encoding (BPE) labels serve as the ground truth in this calculation. The contribution of local loss to the total loss function is regulated by the coefficient $(1 - \alpha)$:

$$\mathcal{L}_{local} = \sum_{X \in \mathcal{D}}\ \sum_{i.. i+k\, \in M} \log P(b_i, \ldots, \ b_{i+k} \ | \ X_{/M})$$
where $b_i$ is the BPE label of $x_i$.

\subsection{Downstream Task Fine-tuning}

A simple multi-layer perceptron (MLP) was exploited to fine-tune the pre-trained encoder for various downstream protein prediction tasks, including classification, regression, and structured prediction problems. 
For sequence-level tasks, e.g., solubility prediction or protein function classification, the \textbf{[CLS]} token embedding can be used to represent the entire input protein sequence. This embedding is fed into the MLP head to make the sequence-level prediction.
For residue-level tasks, e.g., secondary structure prediction or residue property annotation, the output embedding vector corresponding to each residue position can be extracted and fed into the MLP for making per-residue predictions.
Uniquely, in addition to learning single residue representations, our pairwise prediction head allows the model to generate rich pairwise residue-residue representations by definition. This enables fine-tuning the model for tasks involving residue-residue relationships, such as residue contact prediction for modeling tertiary protein structure. The pairwise residue embeddings can be used as input features to the MLP for this task.
Overall, this flexible output representation learning allows our model to be effectively transferred to a diverse array of protein analysis tasks at varying granularities - sequence-level, residue-level, and residue pair-level.

\begin{table*}[!h]
    \centering
    \caption{Comparison of the pure sequence models on the protein function prediction tasks (F1 max / AUPRC). $^*$ indicates structure-based models, whose results are taken from \cite{zhang2022protein}. The highest-performing model is highlighted in bold font, while the second-best performance are indicated by underlining.}
    \vspace{1em}
    \begin{tabular}{c|c|c|c|c}
    \toprule
    Task & GO-MF & GO-BP & GO-CC & EC \\ 
    \hline
    % Metric & \multicolumn{4}{c}{F1 max / AUPRC} \\
    % \midrule
    % ESM-1b$^*$ & - & - & - & 0.657 / 0.639 & 0.452 / 0.332 & 0.477 / 0.324 & 0.864 / 0.889 \\
    GearNet$^*$ & 0.503 / 0.490  & 0.356 / 0.211 & 0.414 / 0.276 & 0.730 / 0.751 \\
    % GearNet-Edge-IEConv$^*$ & - & - & - & 0.581 / - & 0.400 / - & 0.430 / - & 0.810 / - \\
    LM-GVP$^*$ & 0.545 / 0.580 & 0.417 / 0.302 & \textbf{0.527} / \textbf{0.423} & 0.664 / 0.710 \\
    Multiview Contrast$^*$ & 0.654 / 0.596 & \underline{0.490} / 0.292 & 0.488 / 0.336 & \textbf{0.874} / \underline{0.892} \\
    \midrule
    % ESM-1b (650M) & 0.638 & 0.683 & 0.704 & 0.611 / 0.616 & 0.430 / 0.296 & 0.452 / 0.352 & 0.837 / 0.843 \\ 
    ESM2 (650M) & 0.615 / 0.616 & 0.422 / 0.290 &  0.488 / 0.366 & 0.799 / 0.805 \\  
    ESM2 (3B) & \underline{0.659} / \underline{0.647} & 0.476 / \underline{0.341} & 0.497 / \underline{0.417} & 0.863 / 0.876 \\ 
    \ourmodel{} (650M) & 0.649 / 0.645 & 0.471 / 0.330 & \underline{0.516} / 0.395 & 0.855 / 0.884 \\
    \ourmodel{} (3B) & \textbf{0.673} / \textbf{0.657} & \textbf{0.495} / \textbf{0.361} & \underline{0.510} / \underline{0.416} & \underline{0.869} / \textbf{0.893} \\
    \bottomrule
    \end{tabular}
    \label{tab:function_performance}
\end{table*}

\section{Experiment Results}
\label{sec:exp}

\ourmodel{} is pre-trained on the UniRef50 database of clustered protein sequences at 50\% identity. Our model operates on sequences of amino acid residues, comprising over 62 million protein sequences and 17 billion residues from UniRef50. 
For the masked token prediction, we randomly corrupt 30\% of the amino acid residues and then train the model to recover the original residues. This high mask ratio of 30\% allows the model to learn more better representations compared to lower mask ratios.
To efficiently utilize compute resources and enable data-parallel pre-training across multiple GPUs, we develop a novel data packing strategy. Protein sequences are packed into chunks of 8192 tokens, with padding to uniform lengths. This allows batching and parallel processing of sequences during pre-training.
Our masked token prediction and data packing approaches allow scaling pre-training to the full UniRef50 dataset in a computationally efficient manner. We evaluate the pre-trained model on various downstream protein prediction tasks.

% \subsection{Tasks and Metrics}

We compare the performance of different models across various understanding tasks and metrics. The datasets for GO term prediction and EC number prediction are selected from DeepFRI~\cite{gligorijevic2021structure} which are widely used, while the others are primarily adopted from the PEER benchmark~\cite{xu2022peer}. Two model sizes are chosen, namely, 650 million parameters (650M) and 3 billion parameters (3B). The models evaluated include ESM-1b, ESM2, and our model \ourmodel{}. Each model's performance is assessed using specific metrics that are relevant to the particular understanding task. Below, we will analyze these results in detail.
The choice of metric aligns well with the nature of each task, providing a nuanced view of model performance in terms of ranking (Spearman), binary classification accuracy, and multi-class classification accuracy (F1 max, AUPRC).
Our model consistently shows superior performance across nearly all tasks and metrics, highlighting its effectiveness in various biological understanding tasks.

\subsection{\ourmodel{} Exhibits Favorable Scaling Properties}

\begin{figure*}[!h]
    \centering
    \includegraphics[width=0.55\textwidth]{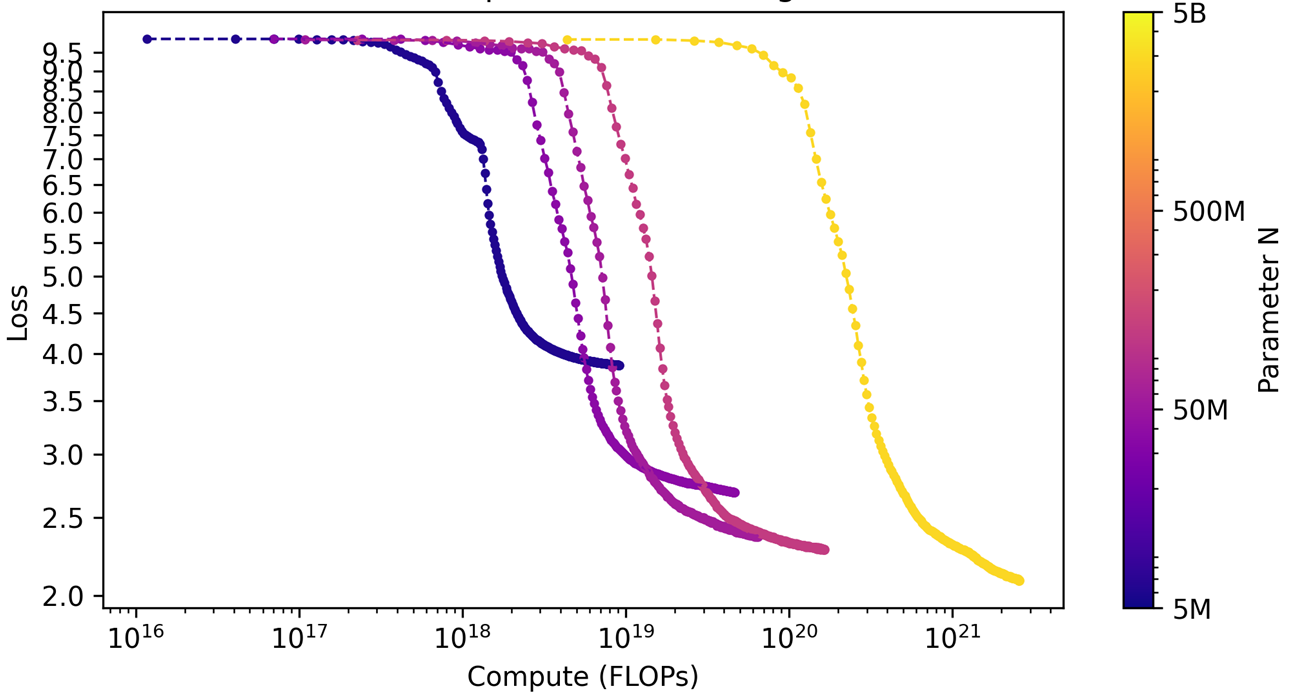}
    \caption{Loss \textit{vs.} Compute during Pre-training}
    \label{fig:scaling_law}
\end{figure*}

During the pre-training phase, we observe a consistent improvement in the model's performance that correlates with increased computational resources, specifically floating-point operations per second (FLOPS), as depicted in Figure~\ref{fig:scaling_law}. In downstream tasks, we note a general enhancement in performance when the number of parameters is augmented from 650 million to 3 billion. This trend holds true for both the ESM2 and our proposed models, indicating that larger models are more adept at capturing intricate patterns and relationships within biological datasets.

\subsection{\ourmodel{} Enables Accurate Protein Function Prediction via Fine-Tuning}

The models are compared on a diverse set of tasks from various benchmark datasets related to protein function annotation, including Gene Ontology (GO) term prediction and Enzyme Commission (EC) number prediction.
Specifically, the tasks include predicting GO terms across the three branches - Molecular Function (GO-MF), Biological Process (GO-BP), and Cellular Component (GO-CC). The EC number prediction task involves assigning EC numbers, which provide a hierarchical classification of enzymes based on the metabolic reactions they catalyze. 
These function prediction tasks use a combination of metrics for evaluation - the maximum F1 score (F1 max) across multiple thresholds, as well as the Area Under the Precision-Recall Curve (AUPRC). F1 max measures the maximum harmonic mean of precision and recall, while AUPRC provides a threshold-independent assessment of the precision-recall trade-off. The results are listed in Table~\ref{tab:function_performance}.

Our model consistently outperforms other state-of-the-art methods across all categories - GO-MF, GO-BP, GO-CC, and EC number prediction - both in terms of the F1 max and AUPRC metrics. Notably, it achieves top scores of 0.855 F1 max and 0.884 AUPRC on the challenging EC number prediction task, surpassing previous models by a significant margin.
Even at the larger 3 billion parameter scale, our model exhibits top performance across all GO categories and EC number prediction. It demonstrates the highest F1 max and AUPRC scores compared to other large language models. This highlights the model's robustness in handling the highly complex and diverse biological data involved in protein function annotation.
Overall, the strong results validate the effectiveness of our self-supervised pre-training approach and pairwise residue modeling for learning powerful protein representations transferable to a wide range of function prediction tasks.

\subsection{\ourmodel{} Enables Data-Driven Protein Fitness Landscape Modeling via Fine-Tuning}

The tasks include quantitative regression tasks like fluorescence prediction and stability prediction for proteins.
These regression tasks use different evaluation metrics compared to the classification tasks. Specifically, Spearman's rank correlation coefficient is employed to assess the correlation between the predicted and ground truth real-valued outputs for fluorescence and stability. The performance are compared in Table~\ref{tab:fitness_performance}.

\begin{table}[!h]
    \centering
    \caption{Comparison of the pure sequence models on the protein fitness prediction tasks (Spearman Correlation).}
    \vspace{1em}
    \begin{tabular}{c|c c}
    \toprule
    Task & Fluorescence & Stability \\ 
    \midrule
    % ESM-1b$^*$ & - & - & - & 0.657 / 0.639 & 0.452 / 0.332 & 0.477 / 0.324 & 0.864 / 0.889 \\
    % ESM-1b (650M) & 0.638 & 0.683 & 0.704 & 0.611 / 0.616 & 0.430 / 0.296 & 0.452 / 0.352 & 0.837 / 0.843 \\ 
    ESM2 (650M) & 0.675 & 0.757  \\  
    ESM2 (3B) & \textbf{0.685} & \underline{0.786} \\ 
    \ourmodel{} (650M) & 0.679 & 0.772  \\
    \ourmodel{} (3B) & \underline{0.683} & \textbf{0.823}  \\
    \bottomrule
    \end{tabular}
    \label{tab:fitness_performance}
\end{table}

On the 650M model scale, our model shows the best overall performance with Spearman correlation coefficients of 0.679 and 0.772 for fluorescence and stability prediction respectively, slightly outperforming the ESM2 and ESM-1b models on these tasks.
At the larger 3 billion parameter scale, our model again excels on the stability prediction task with an impressive Spearman correlation of 0.823, outperforming other large language models. However, for the Fluorescence prediction task, the ESM2 model slightly outperforms ours by a small margin.
The strong results showcase the effectiveness of our model's representations in capturing biophysical properties like stability and fluorescence that are crucial for protein engineering applications. While ESM2 exhibits an edge for the specific case of fluorescence prediction, our model maintains highly competitive performance across most regression tasks in the benchmark.

\subsection{\ourmodel{} Fine-Tuning Enables Accurate Soluble Protein Identification}

The task is solubility prediction, which involves classifying whether a given protein sequence is soluble or insoluble.
For this binary classification task, the accuracy score is used as the evaluation metric. The results are presented in Table~\ref{tab:function_performance}.

\begin{table}[!h]
    \centering
    \caption{Comparison of the pure sequence models on protein solublity prediction (Accuracy).}
    \vspace{1em}
    \begin{tabular}{c|c}
    \toprule
    Task & Solubility \\ 
    \midrule
    % ESM-1b (650M) & 0.638 & 0.683 & 0.704 & 0.611 / 0.616 & 0.430 / 0.296 & 0.452 / 0.352 & 0.837 / 0.843 \\ 
    ESM2 (650M) & 0.747 \\  
    ESM2 (3B) & \underline{0.749} \\ 
    \ourmodel{} (650M) & 0.744 \\
    \ourmodel{} (3B) & \textbf{0.761} \\
    \bottomrule
    \end{tabular}
    \label{tab:solubility_performance}
\end{table}

At the 650M model scale, the ESM2 model exhibits the highest accuracy of 0.747 on solubility prediction, marginally better than our model's accuracy and significantly better than the ESM-1b model.
However, at the larger 3 billion parameter scale, our model takes the lead with an accuracy of 0.761 on the solubility prediction task. This shows that increasing the model's parameter count and capacity can contribute to better understanding and prediction performance on complex biological tasks like solubility.
The results indicate that while ESM2 has a slight edge for solubility at circa 650M parameters, our model is able to better leverage the increased parameter budget to learn more powerful representations. This allows our 3B model to outperform on this challenging task of predicting protein solubility from sequence alone by integrating information across the entire protein chain.
Overall, our model exhibits strong transfer capabilities across diverse tasks in the protein analysis benchmark suite, with the 3B scale achieving state-of-the-art performance on many of the tasks, especially for quantitative regression problems like stability prediction.

\subsection{\ourmodel{} Facilitates Rational Antibody Design via Sequence Completion}

The task of rational antibody CDR-H3 design focuses on constructing the Complementarity-Determining Region H3 (CDR-H3) of an antibody, which plays a critical role in antigen binding due to its high variability and potential for optimization in antibody design. For this study, we employ the RAbD benchmark dataset~\cite{adolf2018rabd}, comprising 60 antibody sequences. The task is framed as designing a CDR-H3 sequence that maximally matches the corresponding residues in the native antibody. 

The evaluation metric is Amino Acid Recovery (AAR), which quantifies the percentage of residues in the generated CDR-H3 sequence that match those in the ground truth sequence. 
The results of the CDR-H3 design task are summarized in Table \ref{tab:rabd_performance}. GNN-based models, such as AR-GNN, RefineGNN~\cite{jin2021refinegnn}, and ABGNN~\cite{gao2023abgnn}, demonstrate moderate performance. Sequence-based models outperform GNN-based approaches, with EnglishBert~\cite{melnyk2023reprogramming} achieving the highest performance. \ourmodel{} delivers comparable results.

\begin{table}[!h]
    \centering
    \caption{Evaluation results on the RAbD benchmark.}
    \vspace{1em}
    % \small
    \begin{tabular}{c|c}
    \toprule
    Model & AAR \\ % $\uparrow$ 
    \midrule
    RAbD & 28.5  \\  
    AR-GNN & 23.8 \\
    Refine-GNN & 35.4 \\
    ABGNN & 39.6 \\
    AbLang & 21.3 \\
    ProtBert & 53.1 \\
    EnglishBert & \textbf{54.9} \\
    ReprogBert & 36.3 \\
    \ourmodel{} (650M) & \underline{54.6}  \\
    \bottomrule
    \end{tabular}
    \label{tab:rabd_performance}
\end{table}

    \section{Discussion}
\label{sec:discussion}

The detailed comparison shows that, across a range of protein understanding tasks, our models, particularly those with 3 billion parameters, demonstrate high effectiveness, largely due to the integrated loss design. These results also suggest that novel approaches are needed for pre-training protein foundation models on pure sequences. Future work should focus on exploring specific features or techniques that contribute to the improved performance of protein foundation models.

	% In the unusual situation where you want a paper to appear in the
	% references without citing it in the main text, use \nocite
	% \nocite{langley00}

    % \bibliography{reference}
    \bibliographystyle{icml2021}
    
\end{document}